\documentclass[aps, reprint, prb, showpacs, longbibliography]{revtex4-1}
\usepackage[version=3]{mhchem}
\usepackage{graphicx}
\usepackage{soul}
\usepackage{floatrow}
\usepackage{amssymb}
\usepackage{amsmath}
\usepackage{commath}
\usepackage{blindtext}
\usepackage{siunitx}
\usepackage{miller}
\usepackage{xcolor}
\usepackage{float}
\usepackage{lipsum}
\usepackage{bm}
\usepackage{soul}
\usepackage{ulem}
\usepackage{physics}
\usepackage{titlesec}
\usepackage{pdfpages}
\usepackage{pgffor}
\titlespacing*{\section}{0pt}{1.2\baselineskip}{\baselineskip}
\titlespacing*{\subsection}{0pt}{1.2\baselineskip}{\baselineskip}

\usepackage{etoolbox} 

\makeatletter
\patchcmd{\@outputpage@head}{\@ifx{\LS@rot\@undefined}{}{\LS@rot}}{}{}{}
\makeatother

\begin{document}

\title{Anisotropic magnetoresistance in spin--orbit semimetal \ce{SrIrO3}}

\author{Dirk\ J. \surname{Groenendijk}}
\email{d.j.groenendijk@tudelft.nl}
\author{Nicola \surname{Manca}}
\author{Joeri \surname{de Bruijckere}}
\author{Ana\ Mafalda\ R.\ V.\ L. \surname{Monteiro}}
\author{Rocco\ \surname{Gaudenzi}}
\author{Herre\ S.\ J. \surname{van der Zant}}
\author{Andrea\ D.\ \surname{Caviglia}}
\email{a.caviglia@tudelft.nl}
\affiliation{Kavli Institute of Nanoscience, Delft University of Technology, Lorentzweg 1, 2628 CJ Delft, Netherlands}

\date{\today}

\begin{abstract}
\ce{SrIrO3}, the three-dimensional member of the Ruddlesden-Popper iridates, is a paramagnetic semimetal characterised by a the delicate interplay between spin--orbit coupling and Coulomb repulsion. In this work, we study the anisotropic magnetoresistance (AMR) of \ce{SrIrO3} thin films, which is closely linked to spin--orbit coupling and probes correlations between electronic transport, magnetic order and orbital states. We show that the low-temperature negative magnetoresistance is anisotropic with respect to the magnetic field orientation, and its angular dependence reveals the appearance of a fourfold symmetric component above a critical magnetic field. We show that this AMR component is of magnetocrystalline origin, and attribute the observed transition to a field-induced magnetic state in \ce{SrIrO3}.
\end{abstract}


\maketitle

\section{Introduction}
5$d$ transition metal oxides feature a rare interplay between Coulomb repulsion $U$, crystal-field $\Delta$ and strong spin--orbit coupling (SOC) that gives rise to novel electronic and magnetic states~\cite{witczak2014correlated, rau2016spin}. A significant body of work has been devoted to the Ruddlesden--Popper series of strontium iridates (\ce{Sr_{$n$+1}Ir_$n$O_{3$n$+1}}) following the discovery of a $j_\mathrm{eff} = 1/2$ Mott state in \ce{Sr2IrO4}~\cite{kim2008novel,kim2009phase}. The dimensionality of these compounds can be tuned by varying $n$, which increases octahedral connectivity and lowers $U$~\cite{kawasaki2016evolution}. The resulting bandwidths have been studied through optical spectroscopy, showing an increase from $0.48\;\mathrm{eV}$ ($n=1$, \ce{Sr2IrO4}) to $1.01\;\mathrm{eV}$ ($n=\infty$, \ce{SrIrO3}), where in the three-dimensional limit a semimetallic state is found~\cite{moon2008dimensionality}. Photoemission and transport studies have shown that the unusual electronic structure of \ce{SrIrO3} (SIO) consists of heavy hole-like and light electron-like bands~\cite{nie2015interplay, liu2016direct, manca2018balanced}. First-principles calculations and diffraction measurements show that these electron-like bands originate from Dirac cones that are gapped due to symmetry breaking in response to strain~\cite{liu2016strain}. This is always the case for epitaxial films, and strain-free SIO can only be studied in polycrystalline form since the single-crystal perovskite phase is thermodynamically unstable~\cite{kim2005metalorganic}. The gapped Dirac semimetallic state has been studied through magnetoresistance (MR) measurements, both in thin films~\cite{wu2013metal,zhang2014sensitively,biswas2014metal,fruchter2016anisotropy,groenendijk2017spin} and in polycrystalline samples~\cite{zhao2008high,fujioka2017correlated}. While the MR in strain-free SIO is 2--3 orders of magnitude larger than in epitaxially strained films, it is qualitatively similar, showing positive, quasilinear behavior.

The proximity of SIO to a metal--insulator phase boundary gives rise to anomalous properties such as non-Fermi liquid behavior and enhanced paramagnetism due to a ferromagnetic instability~\cite{imada1998metal, misawa2007quantum}. Signatures of non-Fermi liquid behavior such as linear resistivity versus temperature~\cite{groenendijk2016epitaxial} and divergent specific heat~\cite{cao2007non} have previously been observed. In ultrathin films, $U$ is further increased by confinement, resulting in an enhancement of spin fluctuations~\cite{groenendijk2017spin}. This brings the system closer to two-dimensional \ce{Sr2IrO4}, in which the magnetic moments display canted in-plane antiferromagnetic order~\cite{crawford1994structural, boseggia2013locking}. The magnetic state of \ce{Sr2IrO4} was studied through anisotropic MR (AMR) measurements, which revealed a field-induced metamagnetic transition from an antiferromagnetic to a weakly ferromagnetic state~\cite{wang2014anisotropic, fina2014anisotropic, lu2018revealing}. Here, we use AMR measurements to study the correlation between electronic transport and magnetic order in ultrathin SIO. We find that the low-temperature negative MR component is anisotropic, and its angular dependence reveals the appearance of a fourfold symmetric component above a critical magnetic field. We attribute this to field-induced magnetic ordering in the SIO film that is inherently close to a ferromagnetic instability.

\section{Results}

\subsection{Sample preparation and experimental setup}
SIO films were grown by pulsed laser deposition on \ce{TiO2}-terminated \ce{SrTiO3}(001) substrates. Hall bars were patterned through Ar etching and subsequent evaporation of \ce{Pd}/\ce{Au} contacts. RHEED oscillations show that both SIO and STO grow in layer-by-layer mode.
Details regarding the growth and fabrication included in the supplementary information and discussed in previous work~\cite{groenendijk2016epitaxial, groenendijk2017spin}. Magnetotransport measurements were performed in a dilution fridge with a base temperature of $70\;\mathrm{mK}$ equipped with a vector magnet and low-noise electronics. The resistance was measured in four-probe configuration with lock-in amplifiers. A Wheatstone bridge circuit was used to measure small resistance variations.

\subsection{Temperature dependence of AMR in \ce{SrIrO3}}

\begin{figure}[b]
\includegraphics[width=\linewidth]{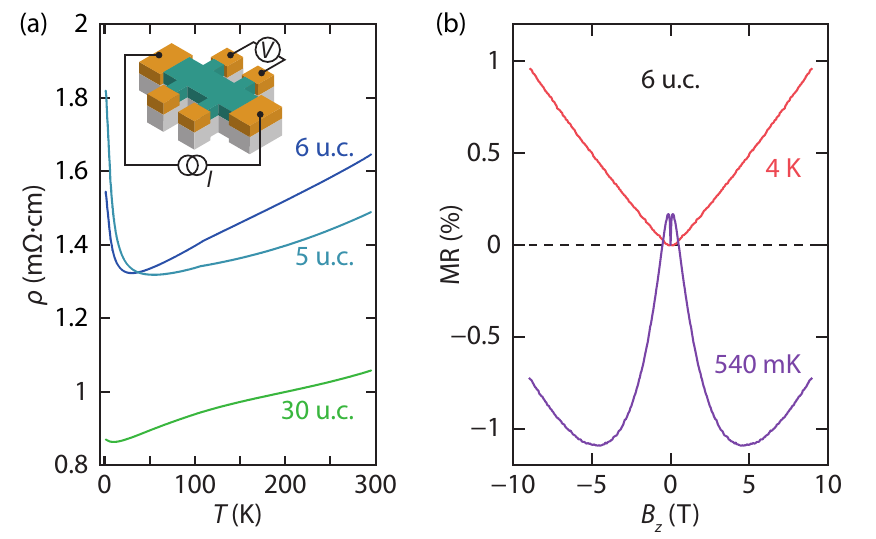}
\caption{\label{Fig1} (a) Resistivity versus temperature of SIO films of different thicknesses grown on STO. (b) Out-of-plane MR of a 6 u.c.~SIO film measured at two different temperatures.}
\end{figure}

Resistivity ($\rho$) versus temperature ($T$) characteristics of three SIO films of different thicknesses are shown in Fig.~\ref{Fig1}a. The films show metallic behavior with an upturn at low temperature similar to previous reports~\cite{biswas2014metal,zhang2014sensitively,groenendijk2016epitaxial,groenendijk2017spin}. The out-of-plane MR of the 6 u.c.~film measured at $T = 4~\mathrm{K}$ and $540~\mathrm{mK}$ is shown in Fig.~\ref{Fig1}b. The MR is quasilinear down to $4\;\mathrm{K}$ and increases in magnitude with decreasing temperature. Its magnitude is approximately 2 orders of magnitude smaller than in polycrystalline samples, where it was attributed to a topological transition of a Dirac node and enhanced paramagnetism~\cite{fujioka2017correlated}. 
At low temperature, a negative MR component appears. For the 6 u.c.~film, this is below $2\;\mathrm{K}$, however this value depends sensitively on the film thickness. MR measurements on the 5 and 30 u.c.~films display similar behavior and are included in the supplementary material. In previous work, we showed that this behavior is governed by a crossover from weak antilocalization to weak localization as the film thickness is reduced~\cite{groenendijk2017spin}.

\begin{figure}[h]
\includegraphics[width=\linewidth]{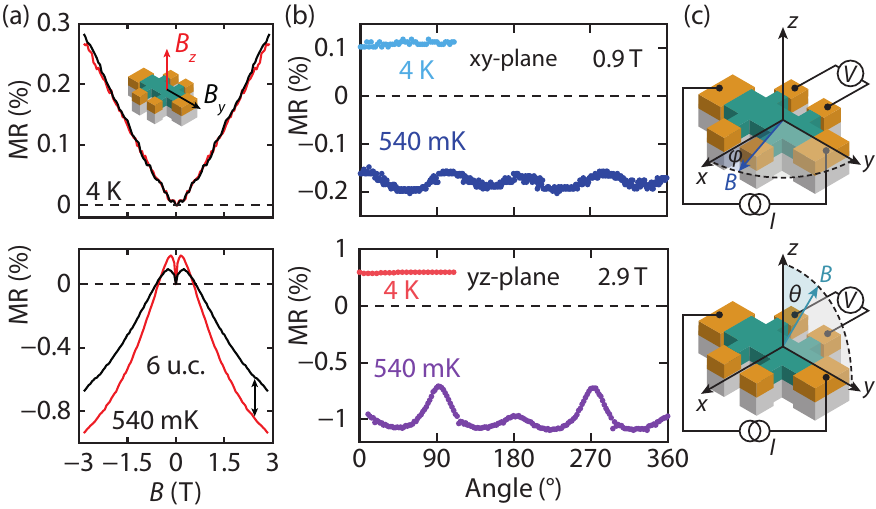}
\caption{\label{Fig2} (a) MR of a 6 u.c.~SIO film measured at $4\;\mathrm{K}$ (top) and $540\;\mathrm{mK}$ (bottom) for magnetic fields applied along $z$ (out-of-plane) and along $y$ (in-plane, $B\parallel I$). (b) Angular dependence of the MR in the $xy$-plane (top) and the $yz$-plane (bottom) at $4\;\mathrm{K}$ and $540\;\mathrm{mK}$. (c) Schematics of the measurement configuration.}
\end{figure}

We study the MR of the 6 u.c.~film further by varying the angle between the magnetic field ($B$) and the film normal. Figure~\ref{Fig2}a shows the MR at $T = 4~\mathrm{K}$ (top) and $540~\mathrm{mK}$ (bottom) for $B$ applied along $z$ and $y$ (see Fig.~\ref{Fig2}c for their definition). At $4~\mathrm{K}$, the two are positive, linear and of equal magnitude, while the negative MR measured at $540~\mathrm{mK}$ shows a pronounced anisotropy. In particular, the magnitude of the negative MR is larger when $B$ is parallel to the film normal. The angular dependence is shown in Fig.~\ref{Fig2}b, where $B$ is rotated with fixed magnitude in the $xy$-plane (top) and in the $yz$-plane (bottom). Interestingly, 4 peaks are observed at $540~\mathrm{mK}$, while there is no measurable anisotropy at $4~\textrm{K}$. Additionally, the MR in the $yz$-plane displays two peaks of larger magnitude, while the peaks heights are equal in the $xy$-plane. 

\subsection{Decoupling of AMR sources}

AMR consists of a noncrystalline and a crystalline component, which have very different microscopic origins. The noncrystalline component depends on the angle between the magnetization ($M$) and current ($I$), reflecting the difference between transport scattering matrix elements for the $I\parallel M$ and $I\perp M$ configurations. The crystalline component, instead, originates from the changes in the equilibrium relativistic electronic structure induced by the rotating magnetization and is thus related to SOC. It manifests itself as a difference between scattering matrix elements for the electrons with momentum parallel and perpendicular to $M$. Owing to the anisotropy of the electronic structure with respect to the magnetization angle, these matrix elements may change when $M$ is rotated~\cite{fina2014anisotropic}.

The two contributions to the AMR can be identified by measuring the magnitude of the MR while rotating $B$ across the different planes defined by our sample geometry (illustrated in Fig.~\ref{Fig2}c). In a rotating magnetic field of strength larger than the coercive field, $M$ follows $B$. This implies that the angle $\theta$ between $M$ and the electrical current $I$ may vary, which, for noncrystalline AMR, results in a signal proportional to $\sin^2(\theta)$. If the magnetic field rotates along in the plane perpendicular to the current ($xz$-plane), the angle between $M$ and $I$ remains constant and the AMR is determined by the varying angle between magnetization and crystal axes. In this way, the crystalline component of the AMR can be isolated.

\begin{figure}[t]
\includegraphics[width=\linewidth]{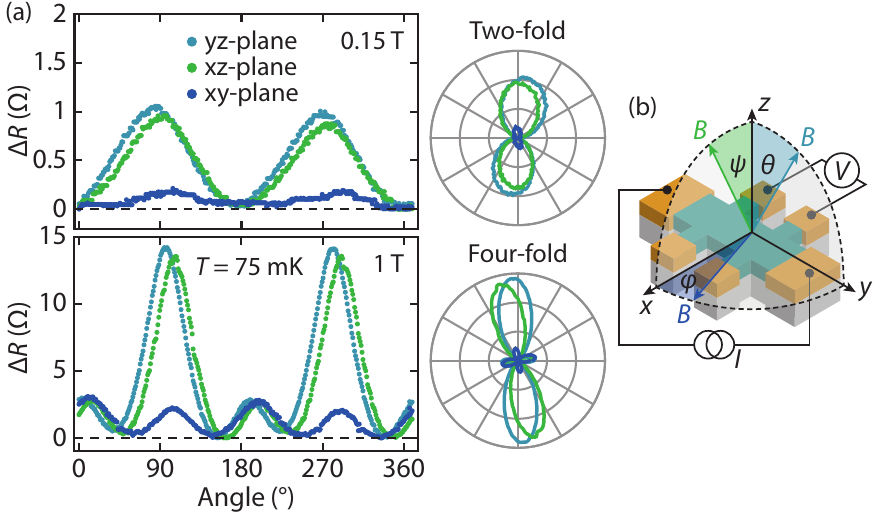}
\caption{\label{Fig3} (a) Angular dependence of the MR in the $xy$-, $xz$- and $yz$-planes measured at $75\;\mathrm{mK}$ with an applied field of $0.15\;\mathrm{T}$ (top) and $1\;\mathrm{T}$ (bottom). On the right, polar plots of the MR are shown. $\Delta R$ is defined as the relative resistance measured during the angular sweep. (b) Schematic of the measurement configuration.}
\end{figure}

Figure~\ref{Fig3} shows the angular dependence of the MR measured while rotating $B$ in the $xy$-, $yz$-, and $xz$-planes. The measurements are performed at the base temperature of the system ($T = 75\;\textrm{mK}$). Since at this temperature the sample resistance ($R$) is large and its variation with magnetic field is small, we use a Wheatstone bridge circuit to accurately measure the change in resistance ($\Delta R$). At $B = 0.15\;\textrm{T}$ [Fig.~\ref{Fig3}a (top)], the symmetry of the AMR is twofold, and the magnitude of $\Delta R$ is nearly equal when rotating the field in the $xz$- and $yz$-planes. The difference (and the small signal in the $xy$-plane) is likely due to a slight misalignment in angle. This indicates that this component [proportional to $\sin^2(\phi)$] is not of noncrystalline origin, as the angle between $I$ and $M$ varies in the $yz$-plane whereas it is always $90^\circ$ in the $xz$-plane. Instead, this component depends on the relative angle between $B$ and the film normal and is also present at low fields. It cannot be attributed to classical MR, as this would provide a positive contribution to the MR when $B$ is perpendicular to the plane. Therefore, this MR most likely originates from the anisotropy of quantum corrections, as the negative MR associated with weak localization is largest when $B$ is along $z$.

At higher fields ($B = 1\;\textrm{T}$, bottom panel), $\Delta R$ increases significantly and  4 peaks of equal magnitude appear in the $xy$-plane. In the $xz$- and $yz$-planes, the peaks at $90^\circ$ and $270^\circ$ increase in magnitude and additional peaks appear at $0^\circ$ and $180^\circ$. The polar plots (right) confirm that the MR in the $xy$-plane is fourfold symmetric, while it shows two large and two small lobes in the $xz$ and $yz$-planes. The magnitude of the fourfold symmetric component [proportional to $\cos^2(2\phi)$] is equal in all planes: this component is thus solely affected by the relative orientation of $B$ and the crystal axes. We note that this AMR component cannot be measured in polycrystalline samples as the contributions from different crystalline domains average out. The sign and symmetry of this component is not compatible with classical MR or weak (anti)localization. The negative MR is largest when the field is oriented at $45^\circ$ with respect to the tetragonal unit cell (see Fig.~\ref{Fig2}b). Since crystalline AMR requires a net magnetization that rotates with respect to the crystal axes, we attribute this to field-induced magnetic ordering in the SIO film. This is consistent with reports of a divergent magnetic susceptibility at low temperatures~\cite{cao2007non} and signatures of a magnetic transition below $2\;\textrm{K}$ in polycrystalline samples~\cite{fujioka2017correlated}.

\subsection{Field-induced magnetic transition}

\begin{figure*}[]
  \floatbox[{\capbeside\thisfloatsetup{capbesideposition={right,center},capbesidewidth=4.5cm}}]{figure}[\FBwidth]
  {\caption{
      (a) Polar plot of the MR in the $yz$-plane measured at $75\;\mathrm{mK}$. The bottom panel shows the relative resistance change along $z$ ($\theta = 0^\circ$) and $y$ ($\theta = 90^\circ$). The fourfold symmetric component appears above $0.2$--$0.3\;\mathrm{T}$.
      (b) Polar plot of the MR in the $xy$-plane. The bottom panel shows the MR along $x$ ($\phi = 0^\circ$) and $y$ ($\phi = 90^\circ$).}
    \label{Fig4}
  }
  {\includegraphics[width=12cm]{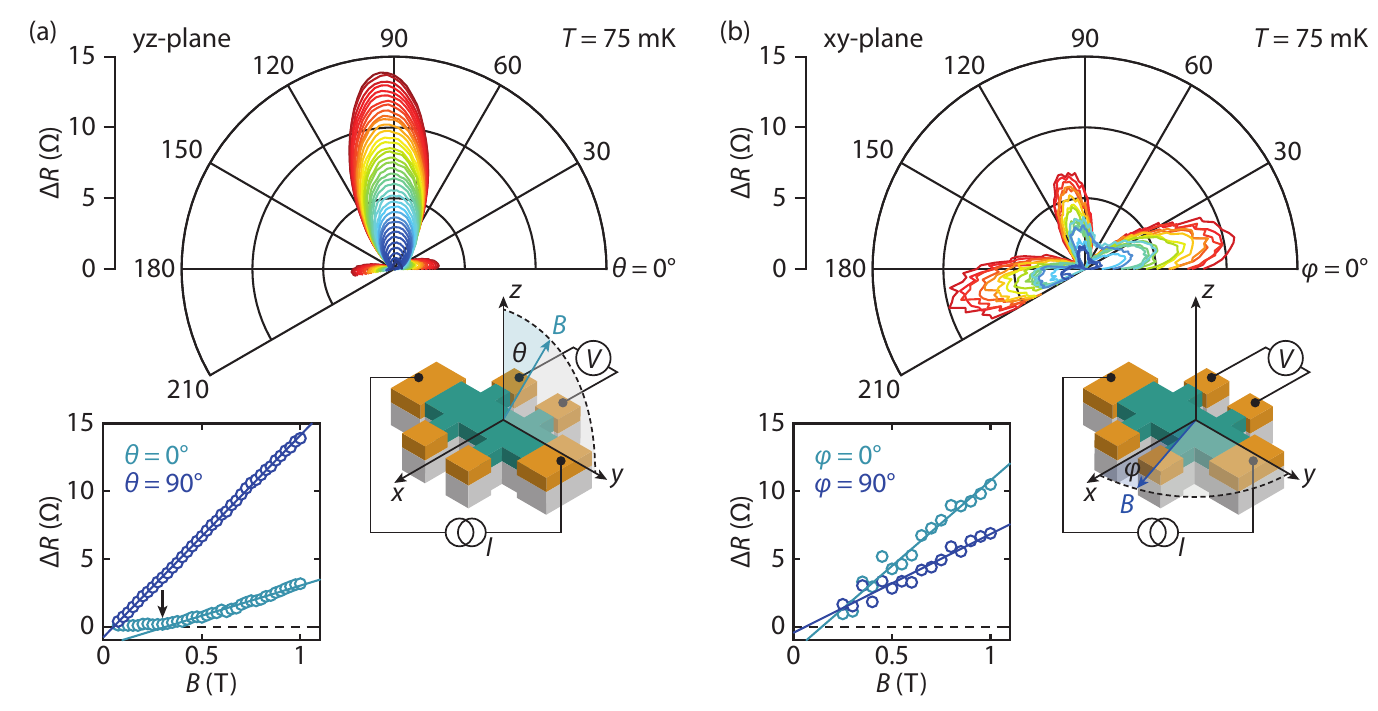}}
\end{figure*}

To determine the field at which the crystalline AMR appears, we measure the angular dependence of the MR for different magnitudes of $B$ as shown in Fig.~\ref{Fig4}a and b. The magnitude of the AMR gradually increases, and the field at which the additional peaks appear can be determined by tracking $\Delta R$ at $\theta$ and $\phi$ = 0 and 90$^\circ$. The bottom panel shows that this occurs at approximately $0.25\;\mathrm{T}$. A field-induced magnetic transition in SIO has previously been inferred from specific heat measurements in monoclinically distorted SIO~\cite{cao2007non}. There, it was suggested that a quantum critical point (QCP) between a non-Fermi liquid and field-induced ferromagnetic state is located at $T=0\;\textrm{K}$ and $\mu_0H = 0.23\;\textrm{T}$. This value corresponds well to the magnetic field at which the fourfold symmetric AMR appears.

Finally, we compare the measured AMR in SIO to \ce{Sr2IrO4}, which is on the other side of the metal--insulator phase boundary. In \ce{Sr2IrO4}, the canting of $j_\textrm{eff} = 1/2$ moments leads to an uncompensated moment within each of the \ce{IrO2} planes, and these moments can be aligned by an external magnetic field, leading to a weakly ferromagnetic state~\cite{wang2014anisotropic, fina2014anisotropic, lu2018revealing}. The magnetic moments are coupled to the octahedral-site rotations by strong spin--orbit coupling, and the AMR can be explained by lattice distortions induced by magnetoelastic coupling. In SIO, such a strong single ion anisotropy is not present. Instead, a enhanced magnetization likely originates from the divergent magnetic susceptibility at low temperature~\cite{cao2007non}. The fourfold symmetry of the AMR is therefore intimately related to the crystal structure, band structure, and orbital symmetry of 5$d$ electrons with strong SOC.

\section{Conclusions}

In conclusion, we showed that the low-temperature, negative MR component in SIO thin films is anisotropic with respect to the magnetic field orientation. The twofold symmetric component, present only when the angle between $B$ and the film normal is varied, is attributed to the anisotropy of weak (anti)localization. A fourfold symmetric component appears as $B$ is increased, and the critical field corresponds well to the quantum critical point previously reported by Cao et al.~\cite{cao2007non}. We attribute this to crystalline AMR resulting from a field-induced magnetic state in ultrathin SIO. Our study underlines the connection between SOC, magnetization and orbital character in strontium iridates. The discovery of a field-induced magnetic state has important implications for interfaces with SIO, where interfacial magnetism, easy-axis reorientation and topological Hall effect have been observed~\cite{nichols2016emerging, yi2016atomic, matsuno2016interface}.

\section{Acknowledgements}
\noindent This work was supported by The Netherlands Organisation for Scientific Research (NWO/OCW) as part of the Frontiers of Nanoscience program (NanoFront), and by the European Research Council under the European Union’s H2020 programme/ERC Grant Agreements No.~[677458] and No.~[731473].

\section{Additional information}
Supplementary information contains details about the growth of SIO and STO thin films and magnetoresistance measurements performed on 5 and 30 u.c. \ce{SrIrO3} films. Correspondence and requests for materials should be addressed to D.J.G.~or A.D.C.

\bibliographystyle{apsrev4-1}
\bibliography{References}

\begin{thebibliography}{30}%
\makeatletter
\providecommand \@ifxundefined [1]{%
 \@ifx{#1\undefined}
}%
\providecommand \@ifnum [1]{%
 \ifnum #1\expandafter \@firstoftwo
 \else \expandafter \@secondoftwo
 \fi
}%
\providecommand \@ifx [1]{%
 \ifx #1\expandafter \@firstoftwo
 \else \expandafter \@secondoftwo
 \fi
}%
\providecommand \natexlab [1]{#1}%
\providecommand \enquote  [1]{``#1''}%
\providecommand \bibnamefont  [1]{#1}%
\providecommand \bibfnamefont [1]{#1}%
\providecommand \citenamefont [1]{#1}%
\providecommand \href@noop [0]{\@secondoftwo}%
\providecommand \href [0]{\begingroup \@sanitize@url \@href}%
\providecommand \@href[1]{\@@startlink{#1}\@@href}%
\providecommand \@@href[1]{\endgroup#1\@@endlink}%
\providecommand \@sanitize@url [0]{\catcode `\\12\catcode `\$12\catcode
  `\&12\catcode `\#12\catcode `\^12\catcode `\_12\catcode `\%12\relax}%
\providecommand \@@startlink[1]{}%
\providecommand \@@endlink[0]{}%
\providecommand \url  [0]{\begingroup\@sanitize@url \@url }%
\providecommand \@url [1]{\endgroup\@href {#1}{\urlprefix }}%
\providecommand \urlprefix  [0]{URL }%
\providecommand \Eprint [0]{\href }%
\providecommand \doibase [0]{http://dx.doi.org/}%
\providecommand \selectlanguage [0]{\@gobble}%
\providecommand \bibinfo  [0]{\@secondoftwo}%
\providecommand \bibfield  [0]{\@secondoftwo}%
\providecommand \translation [1]{[#1]}%
\providecommand \BibitemOpen [0]{}%
\providecommand \bibitemStop [0]{}%
\providecommand \bibitemNoStop [0]{.\EOS\space}%
\providecommand \EOS [0]{\spacefactor3000\relax}%
\providecommand \BibitemShut  [1]{\csname bibitem#1\endcsname}%
\let\auto@bib@innerbib\@empty
\bibitem [{\citenamefont {Witczak-Krempa}\ \emph {et~al.}(2014)\citenamefont
  {Witczak-Krempa}, \citenamefont {Chen}, \citenamefont {Kim},\ and\
  \citenamefont {Balents}}]{witczak2014correlated}%
  \BibitemOpen
  \bibfield  {author} {\bibinfo {author} {\bibfnamefont {W.}~\bibnamefont
  {Witczak-Krempa}}, \bibinfo {author} {\bibfnamefont {G.}~\bibnamefont
  {Chen}}, \bibinfo {author} {\bibfnamefont {Y.~B.}\ \bibnamefont {Kim}}, \
  and\ \bibinfo {author} {\bibfnamefont {L.}~\bibnamefont {Balents}},\
  }\href@noop {} {\bibfield  {journal} {\bibinfo  {journal} {Annu. Rev.
  Condens. Matter Phys.}\ }\textbf {\bibinfo {volume} {5}},\ \bibinfo {pages}
  {57} (\bibinfo {year} {2014})}\BibitemShut {NoStop}%
\bibitem [{\citenamefont {Rau}\ \emph {et~al.}(2016)\citenamefont {Rau},
  \citenamefont {Lee},\ and\ \citenamefont {Kee}}]{rau2016spin}%
  \BibitemOpen
  \bibfield  {author} {\bibinfo {author} {\bibfnamefont {J.~G.}\ \bibnamefont
  {Rau}}, \bibinfo {author} {\bibfnamefont {E.~K.-H.}\ \bibnamefont {Lee}}, \
  and\ \bibinfo {author} {\bibfnamefont {H.-Y.}\ \bibnamefont {Kee}},\
  }\href@noop {} {\bibfield  {journal} {\bibinfo  {journal} {Annual Review of
  Condensed Matter Physics}\ }\textbf {\bibinfo {volume} {7}},\ \bibinfo
  {pages} {195} (\bibinfo {year} {2016})}\BibitemShut {NoStop}%
\bibitem [{\citenamefont {Kim}\ \emph {et~al.}(2008)\citenamefont {Kim},
  \citenamefont {Jin}, \citenamefont {Moon}, \citenamefont {Kim}, \citenamefont
  {Park}, \citenamefont {Leem}, \citenamefont {Yu}, \citenamefont {Noh},
  \citenamefont {Kim}, \citenamefont {Oh} \emph {et~al.}}]{kim2008novel}%
  \BibitemOpen
  \bibfield  {author} {\bibinfo {author} {\bibfnamefont {B.}~\bibnamefont
  {Kim}}, \bibinfo {author} {\bibfnamefont {H.}~\bibnamefont {Jin}}, \bibinfo
  {author} {\bibfnamefont {S.}~\bibnamefont {Moon}}, \bibinfo {author}
  {\bibfnamefont {J.-Y.}\ \bibnamefont {Kim}}, \bibinfo {author} {\bibfnamefont
  {B.-G.}\ \bibnamefont {Park}}, \bibinfo {author} {\bibfnamefont
  {C.}~\bibnamefont {Leem}}, \bibinfo {author} {\bibfnamefont {J.}~\bibnamefont
  {Yu}}, \bibinfo {author} {\bibfnamefont {T.}~\bibnamefont {Noh}}, \bibinfo
  {author} {\bibfnamefont {C.}~\bibnamefont {Kim}}, \bibinfo {author}
  {\bibfnamefont {S.-J.}\ \bibnamefont {Oh}},  \emph {et~al.},\ }\href@noop {}
  {\bibfield  {journal} {\bibinfo  {journal} {Physical Review Letters}\
  }\textbf {\bibinfo {volume} {101}},\ \bibinfo {pages} {076402} (\bibinfo
  {year} {2008})}\BibitemShut {NoStop}%
\bibitem [{\citenamefont {Kim}\ \emph {et~al.}(2009)\citenamefont {Kim},
  \citenamefont {Ohsumi}, \citenamefont {Komesu}, \citenamefont {Sakai},
  \citenamefont {Morita}, \citenamefont {Takagi},\ and\ \citenamefont
  {Arima}}]{kim2009phase}%
  \BibitemOpen
  \bibfield  {author} {\bibinfo {author} {\bibfnamefont {B.}~\bibnamefont
  {Kim}}, \bibinfo {author} {\bibfnamefont {H.}~\bibnamefont {Ohsumi}},
  \bibinfo {author} {\bibfnamefont {T.}~\bibnamefont {Komesu}}, \bibinfo
  {author} {\bibfnamefont {S.}~\bibnamefont {Sakai}}, \bibinfo {author}
  {\bibfnamefont {T.}~\bibnamefont {Morita}}, \bibinfo {author} {\bibfnamefont
  {H.}~\bibnamefont {Takagi}}, \ and\ \bibinfo {author} {\bibfnamefont
  {T.}~\bibnamefont {Arima}},\ }\href@noop {} {\bibfield  {journal} {\bibinfo
  {journal} {Science}\ }\textbf {\bibinfo {volume} {323}},\ \bibinfo {pages}
  {1329} (\bibinfo {year} {2009})}\BibitemShut {NoStop}%
\bibitem [{\citenamefont {Kawasaki}\ \emph {et~al.}(2016)\citenamefont
  {Kawasaki}, \citenamefont {Uchida}, \citenamefont {Paik}, \citenamefont
  {Schlom},\ and\ \citenamefont {Shen}}]{kawasaki2016evolution}%
  \BibitemOpen
  \bibfield  {author} {\bibinfo {author} {\bibfnamefont {J.~K.}\ \bibnamefont
  {Kawasaki}}, \bibinfo {author} {\bibfnamefont {M.}~\bibnamefont {Uchida}},
  \bibinfo {author} {\bibfnamefont {H.}~\bibnamefont {Paik}}, \bibinfo {author}
  {\bibfnamefont {D.~G.}\ \bibnamefont {Schlom}}, \ and\ \bibinfo {author}
  {\bibfnamefont {K.~M.}\ \bibnamefont {Shen}},\ }\href@noop {} {\bibfield
  {journal} {\bibinfo  {journal} {Physical Review B}\ }\textbf {\bibinfo
  {volume} {94}},\ \bibinfo {pages} {121104} (\bibinfo {year}
  {2016})}\BibitemShut {NoStop}%
\bibitem [{\citenamefont {Moon}\ \emph {et~al.}(2008)\citenamefont {Moon},
  \citenamefont {Jin}, \citenamefont {Kim}, \citenamefont {Choi}, \citenamefont
  {Lee}, \citenamefont {Yu}, \citenamefont {Cao}, \citenamefont {Sumi},
  \citenamefont {Funakubo}, \citenamefont {Bernhard} \emph
  {et~al.}}]{moon2008dimensionality}%
  \BibitemOpen
  \bibfield  {author} {\bibinfo {author} {\bibfnamefont {S.}~\bibnamefont
  {Moon}}, \bibinfo {author} {\bibfnamefont {H.}~\bibnamefont {Jin}}, \bibinfo
  {author} {\bibfnamefont {K.~W.}\ \bibnamefont {Kim}}, \bibinfo {author}
  {\bibfnamefont {W.}~\bibnamefont {Choi}}, \bibinfo {author} {\bibfnamefont
  {Y.}~\bibnamefont {Lee}}, \bibinfo {author} {\bibfnamefont {J.}~\bibnamefont
  {Yu}}, \bibinfo {author} {\bibfnamefont {G.}~\bibnamefont {Cao}}, \bibinfo
  {author} {\bibfnamefont {A.}~\bibnamefont {Sumi}}, \bibinfo {author}
  {\bibfnamefont {H.}~\bibnamefont {Funakubo}}, \bibinfo {author}
  {\bibfnamefont {C.}~\bibnamefont {Bernhard}},  \emph {et~al.},\ }\href@noop
  {} {\bibfield  {journal} {\bibinfo  {journal} {Physical Review Letters}\
  }\textbf {\bibinfo {volume} {101}},\ \bibinfo {pages} {226402} (\bibinfo
  {year} {2008})}\BibitemShut {NoStop}%
\bibitem [{\citenamefont {Nie}\ \emph {et~al.}(2015)\citenamefont {Nie},
  \citenamefont {King}, \citenamefont {Kim}, \citenamefont {Uchida},
  \citenamefont {Wei}, \citenamefont {Faeth}, \citenamefont {Ruf},
  \citenamefont {Ruff}, \citenamefont {Xie}, \citenamefont {Pan} \emph
  {et~al.}}]{nie2015interplay}%
  \BibitemOpen
  \bibfield  {author} {\bibinfo {author} {\bibfnamefont {Y.}~\bibnamefont
  {Nie}}, \bibinfo {author} {\bibfnamefont {P.}~\bibnamefont {King}}, \bibinfo
  {author} {\bibfnamefont {C.}~\bibnamefont {Kim}}, \bibinfo {author}
  {\bibfnamefont {M.}~\bibnamefont {Uchida}}, \bibinfo {author} {\bibfnamefont
  {H.}~\bibnamefont {Wei}}, \bibinfo {author} {\bibfnamefont {B.}~\bibnamefont
  {Faeth}}, \bibinfo {author} {\bibfnamefont {J.}~\bibnamefont {Ruf}}, \bibinfo
  {author} {\bibfnamefont {J.}~\bibnamefont {Ruff}}, \bibinfo {author}
  {\bibfnamefont {L.}~\bibnamefont {Xie}}, \bibinfo {author} {\bibfnamefont
  {X.}~\bibnamefont {Pan}},  \emph {et~al.},\ }\href@noop {} {\bibfield
  {journal} {\bibinfo  {journal} {Physical Review Letters}\ }\textbf {\bibinfo
  {volume} {114}},\ \bibinfo {pages} {016401} (\bibinfo {year}
  {2015})}\BibitemShut {NoStop}%
\bibitem [{\citenamefont {Liu}\ \emph {et~al.}(2016{\natexlab{a}})\citenamefont
  {Liu}, \citenamefont {Li}, \citenamefont {Li}, \citenamefont {Liu},
  \citenamefont {Li}, \citenamefont {Yang}, \citenamefont {Yao}, \citenamefont
  {Fan}, \citenamefont {Wan}, \citenamefont {Wang} \emph
  {et~al.}}]{liu2016direct}%
  \BibitemOpen
  \bibfield  {author} {\bibinfo {author} {\bibfnamefont {Z.}~\bibnamefont
  {Liu}}, \bibinfo {author} {\bibfnamefont {M.}~\bibnamefont {Li}}, \bibinfo
  {author} {\bibfnamefont {Q.}~\bibnamefont {Li}}, \bibinfo {author}
  {\bibfnamefont {J.}~\bibnamefont {Liu}}, \bibinfo {author} {\bibfnamefont
  {W.}~\bibnamefont {Li}}, \bibinfo {author} {\bibfnamefont {H.}~\bibnamefont
  {Yang}}, \bibinfo {author} {\bibfnamefont {Q.}~\bibnamefont {Yao}}, \bibinfo
  {author} {\bibfnamefont {C.}~\bibnamefont {Fan}}, \bibinfo {author}
  {\bibfnamefont {X.}~\bibnamefont {Wan}}, \bibinfo {author} {\bibfnamefont
  {Z.}~\bibnamefont {Wang}},  \emph {et~al.},\ }\href@noop {} {\bibfield
  {journal} {\bibinfo  {journal} {Scientific Reports}\ }\textbf {\bibinfo
  {volume} {6}},\ \bibinfo {pages} {30309} (\bibinfo {year}
  {2016}{\natexlab{a}})}\BibitemShut {NoStop}%
\bibitem [{\citenamefont {Manca}\ \emph {et~al.}(2018)\citenamefont {Manca},
  \citenamefont {Groenendijk}, \citenamefont {Pallecchi}, \citenamefont
  {Autieri}, \citenamefont {Tang}, \citenamefont {Telesio}, \citenamefont
  {Mattoni}, \citenamefont {McCollam}, \citenamefont {Picozzi},\ and\
  \citenamefont {Caviglia}}]{manca2018balanced}%
  \BibitemOpen
  \bibfield  {author} {\bibinfo {author} {\bibfnamefont {N.}~\bibnamefont
  {Manca}}, \bibinfo {author} {\bibfnamefont {D.~J.}\ \bibnamefont
  {Groenendijk}}, \bibinfo {author} {\bibfnamefont {I.}~\bibnamefont
  {Pallecchi}}, \bibinfo {author} {\bibfnamefont {C.}~\bibnamefont {Autieri}},
  \bibinfo {author} {\bibfnamefont {L.~M.~K.}\ \bibnamefont {Tang}}, \bibinfo
  {author} {\bibfnamefont {F.}~\bibnamefont {Telesio}}, \bibinfo {author}
  {\bibfnamefont {G.}~\bibnamefont {Mattoni}}, \bibinfo {author} {\bibfnamefont
  {A.}~\bibnamefont {McCollam}}, \bibinfo {author} {\bibfnamefont
  {S.}~\bibnamefont {Picozzi}}, \ and\ \bibinfo {author} {\bibfnamefont
  {A.~D.}\ \bibnamefont {Caviglia}},\ }\href {\doibase
  10.1103/PhysRevB.97.081105} {\bibfield  {journal} {\bibinfo  {journal} {Phys.
  Rev. B}\ }\textbf {\bibinfo {volume} {97}},\ \bibinfo {pages} {081105}
  (\bibinfo {year} {2018})}\BibitemShut {NoStop}%
\bibitem [{\citenamefont {Liu}\ \emph {et~al.}(2016{\natexlab{b}})\citenamefont
  {Liu}, \citenamefont {Kriegner}, \citenamefont {Horak}, \citenamefont
  {Puggioni}, \citenamefont {Serrao}, \citenamefont {Chen}, \citenamefont {Yi},
  \citenamefont {Frontera}, \citenamefont {Holy}, \citenamefont {Vishwanath}
  \emph {et~al.}}]{liu2016strain}%
  \BibitemOpen
  \bibfield  {author} {\bibinfo {author} {\bibfnamefont {J.}~\bibnamefont
  {Liu}}, \bibinfo {author} {\bibfnamefont {D.}~\bibnamefont {Kriegner}},
  \bibinfo {author} {\bibfnamefont {L.}~\bibnamefont {Horak}}, \bibinfo
  {author} {\bibfnamefont {D.}~\bibnamefont {Puggioni}}, \bibinfo {author}
  {\bibfnamefont {C.~R.}\ \bibnamefont {Serrao}}, \bibinfo {author}
  {\bibfnamefont {R.}~\bibnamefont {Chen}}, \bibinfo {author} {\bibfnamefont
  {D.}~\bibnamefont {Yi}}, \bibinfo {author} {\bibfnamefont {C.}~\bibnamefont
  {Frontera}}, \bibinfo {author} {\bibfnamefont {V.}~\bibnamefont {Holy}},
  \bibinfo {author} {\bibfnamefont {A.}~\bibnamefont {Vishwanath}},  \emph
  {et~al.},\ }\href@noop {} {\bibfield  {journal} {\bibinfo  {journal}
  {Physical Review B}\ }\textbf {\bibinfo {volume} {93}},\ \bibinfo {pages}
  {085118} (\bibinfo {year} {2016}{\natexlab{b}})}\BibitemShut {NoStop}%
\bibitem [{\citenamefont {Kim}\ \emph {et~al.}(2005)\citenamefont {Kim},
  \citenamefont {Sumi}, \citenamefont {Takahashi}, \citenamefont {Yokoyama},
  \citenamefont {Ito}, \citenamefont {Watanabe}, \citenamefont {Akiyama},
  \citenamefont {Kaneko}, \citenamefont {Saito},\ and\ \citenamefont
  {Funakubo}}]{kim2005metalorganic}%
  \BibitemOpen
  \bibfield  {author} {\bibinfo {author} {\bibfnamefont {Y.~K.}\ \bibnamefont
  {Kim}}, \bibinfo {author} {\bibfnamefont {A.}~\bibnamefont {Sumi}}, \bibinfo
  {author} {\bibfnamefont {K.}~\bibnamefont {Takahashi}}, \bibinfo {author}
  {\bibfnamefont {S.}~\bibnamefont {Yokoyama}}, \bibinfo {author}
  {\bibfnamefont {S.}~\bibnamefont {Ito}}, \bibinfo {author} {\bibfnamefont
  {T.}~\bibnamefont {Watanabe}}, \bibinfo {author} {\bibfnamefont
  {K.}~\bibnamefont {Akiyama}}, \bibinfo {author} {\bibfnamefont
  {S.}~\bibnamefont {Kaneko}}, \bibinfo {author} {\bibfnamefont
  {K.}~\bibnamefont {Saito}}, \ and\ \bibinfo {author} {\bibfnamefont
  {H.}~\bibnamefont {Funakubo}},\ }\href@noop {} {\bibfield  {journal}
  {\bibinfo  {journal} {Japanese journal of applied physics}\ }\textbf
  {\bibinfo {volume} {45}},\ \bibinfo {pages} {L36} (\bibinfo {year}
  {2005})}\BibitemShut {NoStop}%
\bibitem [{\citenamefont {Wu}\ \emph {et~al.}(2013)\citenamefont {Wu},
  \citenamefont {Zhou}, \citenamefont {Zhang}, \citenamefont {Chen},
  \citenamefont {Zhang}, \citenamefont {Gu}, \citenamefont {Yao},\ and\
  \citenamefont {Chen}}]{wu2013metal}%
  \BibitemOpen
  \bibfield  {author} {\bibinfo {author} {\bibfnamefont {F.-X.}\ \bibnamefont
  {Wu}}, \bibinfo {author} {\bibfnamefont {J.}~\bibnamefont {Zhou}}, \bibinfo
  {author} {\bibfnamefont {L.}~\bibnamefont {Zhang}}, \bibinfo {author}
  {\bibfnamefont {Y.}~\bibnamefont {Chen}}, \bibinfo {author} {\bibfnamefont
  {S.-T.}\ \bibnamefont {Zhang}}, \bibinfo {author} {\bibfnamefont {Z.-B.}\
  \bibnamefont {Gu}}, \bibinfo {author} {\bibfnamefont {S.-H.}\ \bibnamefont
  {Yao}}, \ and\ \bibinfo {author} {\bibfnamefont {Y.-F.}\ \bibnamefont
  {Chen}},\ }\href@noop {} {\bibfield  {journal} {\bibinfo  {journal} {Journal
  of Physics: Condensed Matter}\ }\textbf {\bibinfo {volume} {25}},\ \bibinfo
  {pages} {125604} (\bibinfo {year} {2013})}\BibitemShut {NoStop}%
\bibitem [{\citenamefont {Zhang}\ \emph {et~al.}(2014)\citenamefont {Zhang},
  \citenamefont {Chen}, \citenamefont {Zhang}, \citenamefont {Zhou},
  \citenamefont {Zhang}, \citenamefont {Gu}, \citenamefont {Yao},\ and\
  \citenamefont {Chen}}]{zhang2014sensitively}%
  \BibitemOpen
  \bibfield  {author} {\bibinfo {author} {\bibfnamefont {L.}~\bibnamefont
  {Zhang}}, \bibinfo {author} {\bibfnamefont {Y.}~\bibnamefont {Chen}},
  \bibinfo {author} {\bibfnamefont {B.}~\bibnamefont {Zhang}}, \bibinfo
  {author} {\bibfnamefont {J.}~\bibnamefont {Zhou}}, \bibinfo {author}
  {\bibfnamefont {S.}~\bibnamefont {Zhang}}, \bibinfo {author} {\bibfnamefont
  {Z.}~\bibnamefont {Gu}}, \bibinfo {author} {\bibfnamefont {S.}~\bibnamefont
  {Yao}}, \ and\ \bibinfo {author} {\bibfnamefont {Y.}~\bibnamefont {Chen}},\
  }\href@noop {} {\bibfield  {journal} {\bibinfo  {journal} {Journal of the
  Physical Society of Japan}\ }\textbf {\bibinfo {volume} {83}},\ \bibinfo
  {pages} {054707} (\bibinfo {year} {2014})}\BibitemShut {NoStop}%
\bibitem [{\citenamefont {Biswas}\ \emph {et~al.}(2014)\citenamefont {Biswas},
  \citenamefont {Kim},\ and\ \citenamefont {Jeong}}]{biswas2014metal}%
  \BibitemOpen
  \bibfield  {author} {\bibinfo {author} {\bibfnamefont {A.}~\bibnamefont
  {Biswas}}, \bibinfo {author} {\bibfnamefont {K.-S.}\ \bibnamefont {Kim}}, \
  and\ \bibinfo {author} {\bibfnamefont {Y.~H.}\ \bibnamefont {Jeong}},\
  }\href@noop {} {\bibfield  {journal} {\bibinfo  {journal} {Journal of Applied
  Physics}\ }\textbf {\bibinfo {volume} {116}},\ \bibinfo {pages} {213704}
  (\bibinfo {year} {2014})}\BibitemShut {NoStop}%
\bibitem [{\citenamefont {Fruchter}\ \emph {et~al.}(2016)\citenamefont
  {Fruchter}, \citenamefont {Schneegans},\ and\ \citenamefont
  {Li}}]{fruchter2016anisotropy}%
  \BibitemOpen
  \bibfield  {author} {\bibinfo {author} {\bibfnamefont {L.}~\bibnamefont
  {Fruchter}}, \bibinfo {author} {\bibfnamefont {O.}~\bibnamefont
  {Schneegans}}, \ and\ \bibinfo {author} {\bibfnamefont {Z.}~\bibnamefont
  {Li}},\ }\href@noop {} {\bibfield  {journal} {\bibinfo  {journal} {Journal of
  Applied Physics}\ }\textbf {\bibinfo {volume} {120}},\ \bibinfo {pages}
  {075307} (\bibinfo {year} {2016})}\BibitemShut {NoStop}%
\bibitem [{\citenamefont {Groenendijk}\ \emph {et~al.}(2017)\citenamefont
  {Groenendijk}, \citenamefont {Autieri}, \citenamefont {Girovsky},
  \citenamefont {Martinez-Velarte}, \citenamefont {Manca}, \citenamefont
  {Mattoni}, \citenamefont {Monteiro}, \citenamefont {Gauquelin}, \citenamefont
  {Verbeeck}, \citenamefont {Otte} \emph {et~al.}}]{groenendijk2017spin}%
  \BibitemOpen
  \bibfield  {author} {\bibinfo {author} {\bibfnamefont {D.}~\bibnamefont
  {Groenendijk}}, \bibinfo {author} {\bibfnamefont {C.}~\bibnamefont
  {Autieri}}, \bibinfo {author} {\bibfnamefont {J.}~\bibnamefont {Girovsky}},
  \bibinfo {author} {\bibfnamefont {M.~C.}\ \bibnamefont {Martinez-Velarte}},
  \bibinfo {author} {\bibfnamefont {N.}~\bibnamefont {Manca}}, \bibinfo
  {author} {\bibfnamefont {G.}~\bibnamefont {Mattoni}}, \bibinfo {author}
  {\bibfnamefont {A.}~\bibnamefont {Monteiro}}, \bibinfo {author}
  {\bibfnamefont {N.}~\bibnamefont {Gauquelin}}, \bibinfo {author}
  {\bibfnamefont {J.}~\bibnamefont {Verbeeck}}, \bibinfo {author}
  {\bibfnamefont {A.}~\bibnamefont {Otte}},  \emph {et~al.},\ }\href@noop {}
  {\bibfield  {journal} {\bibinfo  {journal} {Physical Review Letters}\
  }\textbf {\bibinfo {volume} {119}},\ \bibinfo {pages} {256403} (\bibinfo
  {year} {2017})}\BibitemShut {NoStop}%
\bibitem [{\citenamefont {Zhao}\ \emph {et~al.}(2008)\citenamefont {Zhao},
  \citenamefont {Yang}, \citenamefont {Yu}, \citenamefont {Li}, \citenamefont
  {Yu}, \citenamefont {Fang}, \citenamefont {Chen},\ and\ \citenamefont
  {Jin}}]{zhao2008high}%
  \BibitemOpen
  \bibfield  {author} {\bibinfo {author} {\bibfnamefont {J.}~\bibnamefont
  {Zhao}}, \bibinfo {author} {\bibfnamefont {L.}~\bibnamefont {Yang}}, \bibinfo
  {author} {\bibfnamefont {Y.}~\bibnamefont {Yu}}, \bibinfo {author}
  {\bibfnamefont {F.}~\bibnamefont {Li}}, \bibinfo {author} {\bibfnamefont
  {R.}~\bibnamefont {Yu}}, \bibinfo {author} {\bibfnamefont {Z.}~\bibnamefont
  {Fang}}, \bibinfo {author} {\bibfnamefont {L.}~\bibnamefont {Chen}}, \ and\
  \bibinfo {author} {\bibfnamefont {C.}~\bibnamefont {Jin}},\ }\href@noop {}
  {\bibfield  {journal} {\bibinfo  {journal} {Journal of Applied Physics}\
  }\textbf {\bibinfo {volume} {103}},\ \bibinfo {pages} {103706} (\bibinfo
  {year} {2008})}\BibitemShut {NoStop}%
\bibitem [{\citenamefont {Fujioka}\ \emph {et~al.}(2017)\citenamefont
  {Fujioka}, \citenamefont {Okawa}, \citenamefont {Yamamoto},\ and\
  \citenamefont {Tokura}}]{fujioka2017correlated}%
  \BibitemOpen
  \bibfield  {author} {\bibinfo {author} {\bibfnamefont {J.}~\bibnamefont
  {Fujioka}}, \bibinfo {author} {\bibfnamefont {T.}~\bibnamefont {Okawa}},
  \bibinfo {author} {\bibfnamefont {A.}~\bibnamefont {Yamamoto}}, \ and\
  \bibinfo {author} {\bibfnamefont {Y.}~\bibnamefont {Tokura}},\ }\href@noop {}
  {\bibfield  {journal} {\bibinfo  {journal} {Physical Review B}\ }\textbf
  {\bibinfo {volume} {95}},\ \bibinfo {pages} {121102} (\bibinfo {year}
  {2017})}\BibitemShut {NoStop}%
\bibitem [{\citenamefont {Imada}\ \emph {et~al.}(1998)\citenamefont {Imada},
  \citenamefont {Fujimori},\ and\ \citenamefont {Tokura}}]{imada1998metal}%
  \BibitemOpen
  \bibfield  {author} {\bibinfo {author} {\bibfnamefont {M.}~\bibnamefont
  {Imada}}, \bibinfo {author} {\bibfnamefont {A.}~\bibnamefont {Fujimori}}, \
  and\ \bibinfo {author} {\bibfnamefont {Y.}~\bibnamefont {Tokura}},\
  }\href@noop {} {\bibfield  {journal} {\bibinfo  {journal} {Reviews of Modern
  Physics}\ }\textbf {\bibinfo {volume} {70}},\ \bibinfo {pages} {1039}
  (\bibinfo {year} {1998})}\BibitemShut {NoStop}%
\bibitem [{\citenamefont {Misawa}\ and\ \citenamefont
  {Imada}(2007)}]{misawa2007quantum}%
  \BibitemOpen
  \bibfield  {author} {\bibinfo {author} {\bibfnamefont {T.}~\bibnamefont
  {Misawa}}\ and\ \bibinfo {author} {\bibfnamefont {M.}~\bibnamefont {Imada}},\
  }\href@noop {} {\bibfield  {journal} {\bibinfo  {journal} {Physical Review
  B}\ }\textbf {\bibinfo {volume} {75}},\ \bibinfo {pages} {115121} (\bibinfo
  {year} {2007})}\BibitemShut {NoStop}%
\bibitem [{\citenamefont {Groenendijk}\ \emph {et~al.}(2016)\citenamefont
  {Groenendijk}, \citenamefont {Manca}, \citenamefont {Mattoni}, \citenamefont
  {Kootstra}, \citenamefont {Gariglio}, \citenamefont {Huang}, \citenamefont
  {van Heumen},\ and\ \citenamefont {Caviglia}}]{groenendijk2016epitaxial}%
  \BibitemOpen
  \bibfield  {author} {\bibinfo {author} {\bibfnamefont {D.}~\bibnamefont
  {Groenendijk}}, \bibinfo {author} {\bibfnamefont {N.}~\bibnamefont {Manca}},
  \bibinfo {author} {\bibfnamefont {G.}~\bibnamefont {Mattoni}}, \bibinfo
  {author} {\bibfnamefont {L.}~\bibnamefont {Kootstra}}, \bibinfo {author}
  {\bibfnamefont {S.}~\bibnamefont {Gariglio}}, \bibinfo {author}
  {\bibfnamefont {Y.}~\bibnamefont {Huang}}, \bibinfo {author} {\bibfnamefont
  {E.}~\bibnamefont {van Heumen}}, \ and\ \bibinfo {author} {\bibfnamefont
  {A.}~\bibnamefont {Caviglia}},\ }\href@noop {} {\bibfield  {journal}
  {\bibinfo  {journal} {Applied Physics Letters}\ }\textbf {\bibinfo {volume}
  {109}},\ \bibinfo {pages} {041906} (\bibinfo {year} {2016})}\BibitemShut
  {NoStop}%
\bibitem [{\citenamefont {Cao}\ \emph {et~al.}(2007)\citenamefont {Cao},
  \citenamefont {Durairaj}, \citenamefont {Chikara}, \citenamefont {DeLong},
  \citenamefont {Parkin},\ and\ \citenamefont {Schlottmann}}]{cao2007non}%
  \BibitemOpen
  \bibfield  {author} {\bibinfo {author} {\bibfnamefont {G.}~\bibnamefont
  {Cao}}, \bibinfo {author} {\bibfnamefont {V.}~\bibnamefont {Durairaj}},
  \bibinfo {author} {\bibfnamefont {S.}~\bibnamefont {Chikara}}, \bibinfo
  {author} {\bibfnamefont {L.}~\bibnamefont {DeLong}}, \bibinfo {author}
  {\bibfnamefont {S.}~\bibnamefont {Parkin}}, \ and\ \bibinfo {author}
  {\bibfnamefont {P.}~\bibnamefont {Schlottmann}},\ }\href@noop {} {\bibfield
  {journal} {\bibinfo  {journal} {Physical Review B}\ }\textbf {\bibinfo
  {volume} {76}},\ \bibinfo {pages} {100402} (\bibinfo {year}
  {2007})}\BibitemShut {NoStop}%
\bibitem [{\citenamefont {Crawford}\ \emph {et~al.}(1994)\citenamefont
  {Crawford}, \citenamefont {Subramanian}, \citenamefont {Harlow},
  \citenamefont {Fernandez-Baca}, \citenamefont {Wang},\ and\ \citenamefont
  {Johnston}}]{crawford1994structural}%
  \BibitemOpen
  \bibfield  {author} {\bibinfo {author} {\bibfnamefont {M.}~\bibnamefont
  {Crawford}}, \bibinfo {author} {\bibfnamefont {M.}~\bibnamefont
  {Subramanian}}, \bibinfo {author} {\bibfnamefont {R.}~\bibnamefont {Harlow}},
  \bibinfo {author} {\bibfnamefont {J.}~\bibnamefont {Fernandez-Baca}},
  \bibinfo {author} {\bibfnamefont {Z.}~\bibnamefont {Wang}}, \ and\ \bibinfo
  {author} {\bibfnamefont {D.}~\bibnamefont {Johnston}},\ }\href@noop {}
  {\bibfield  {journal} {\bibinfo  {journal} {Physical Review B}\ }\textbf
  {\bibinfo {volume} {49}},\ \bibinfo {pages} {9198} (\bibinfo {year}
  {1994})}\BibitemShut {NoStop}%
\bibitem [{\citenamefont {Boseggia}\ \emph {et~al.}(2013)\citenamefont
  {Boseggia}, \citenamefont {Walker}, \citenamefont {Vale}, \citenamefont
  {Springell}, \citenamefont {Feng}, \citenamefont {Perry}, \citenamefont
  {Sala}, \citenamefont {R{\o}nnow}, \citenamefont {Collins},\ and\
  \citenamefont {McMorrow}}]{boseggia2013locking}%
  \BibitemOpen
  \bibfield  {author} {\bibinfo {author} {\bibfnamefont {S.}~\bibnamefont
  {Boseggia}}, \bibinfo {author} {\bibfnamefont {H.}~\bibnamefont {Walker}},
  \bibinfo {author} {\bibfnamefont {J.}~\bibnamefont {Vale}}, \bibinfo {author}
  {\bibfnamefont {R.}~\bibnamefont {Springell}}, \bibinfo {author}
  {\bibfnamefont {Z.}~\bibnamefont {Feng}}, \bibinfo {author} {\bibfnamefont
  {R.}~\bibnamefont {Perry}}, \bibinfo {author} {\bibfnamefont {M.~M.}\
  \bibnamefont {Sala}}, \bibinfo {author} {\bibfnamefont {H.~M.}\ \bibnamefont
  {R{\o}nnow}}, \bibinfo {author} {\bibfnamefont {S.}~\bibnamefont {Collins}},
  \ and\ \bibinfo {author} {\bibfnamefont {D.~F.}\ \bibnamefont {McMorrow}},\
  }\href@noop {} {\bibfield  {journal} {\bibinfo  {journal} {Journal of
  Physics: Condensed Matter}\ }\textbf {\bibinfo {volume} {25}},\ \bibinfo
  {pages} {422202} (\bibinfo {year} {2013})}\BibitemShut {NoStop}%
\bibitem [{\citenamefont {Wang}\ \emph {et~al.}(2014)\citenamefont {Wang},
  \citenamefont {Seinige}, \citenamefont {Cao}, \citenamefont {Zhou},
  \citenamefont {Goodenough},\ and\ \citenamefont
  {Tsoi}}]{wang2014anisotropic}%
  \BibitemOpen
  \bibfield  {author} {\bibinfo {author} {\bibfnamefont {C.}~\bibnamefont
  {Wang}}, \bibinfo {author} {\bibfnamefont {H.}~\bibnamefont {Seinige}},
  \bibinfo {author} {\bibfnamefont {G.}~\bibnamefont {Cao}}, \bibinfo {author}
  {\bibfnamefont {J.-S.}\ \bibnamefont {Zhou}}, \bibinfo {author}
  {\bibfnamefont {J.~B.}\ \bibnamefont {Goodenough}}, \ and\ \bibinfo {author}
  {\bibfnamefont {M.}~\bibnamefont {Tsoi}},\ }\href@noop {} {\bibfield
  {journal} {\bibinfo  {journal} {Physical Review X}\ }\textbf {\bibinfo
  {volume} {4}},\ \bibinfo {pages} {041034} (\bibinfo {year}
  {2014})}\BibitemShut {NoStop}%
\bibitem [{\citenamefont {Fina}\ \emph {et~al.}(2014)\citenamefont {Fina},
  \citenamefont {Marti}, \citenamefont {Yi}, \citenamefont {Liu}, \citenamefont
  {Chu}, \citenamefont {Rayan-Serrao}, \citenamefont {Suresha}, \citenamefont
  {Shick}, \citenamefont {{\v{Z}}elezn{\`y}}, \citenamefont {Jungwirth} \emph
  {et~al.}}]{fina2014anisotropic}%
  \BibitemOpen
  \bibfield  {author} {\bibinfo {author} {\bibfnamefont {I.}~\bibnamefont
  {Fina}}, \bibinfo {author} {\bibfnamefont {X.}~\bibnamefont {Marti}},
  \bibinfo {author} {\bibfnamefont {D.}~\bibnamefont {Yi}}, \bibinfo {author}
  {\bibfnamefont {J.}~\bibnamefont {Liu}}, \bibinfo {author} {\bibfnamefont
  {J.-H.}\ \bibnamefont {Chu}}, \bibinfo {author} {\bibfnamefont
  {C.}~\bibnamefont {Rayan-Serrao}}, \bibinfo {author} {\bibfnamefont
  {S.}~\bibnamefont {Suresha}}, \bibinfo {author} {\bibfnamefont
  {A.}~\bibnamefont {Shick}}, \bibinfo {author} {\bibfnamefont
  {J.}~\bibnamefont {{\v{Z}}elezn{\`y}}}, \bibinfo {author} {\bibfnamefont
  {T.}~\bibnamefont {Jungwirth}},  \emph {et~al.},\ }\href@noop {} {\bibfield
  {journal} {\bibinfo  {journal} {Nature communications}\ }\textbf {\bibinfo
  {volume} {5}},\ \bibinfo {pages} {4671} (\bibinfo {year} {2014})}\BibitemShut
  {NoStop}%
\bibitem [{\citenamefont {Lu}\ \emph {et~al.}(2018)\citenamefont {Lu},
  \citenamefont {Gao}, \citenamefont {Wang}, \citenamefont {Wang},
  \citenamefont {Yuan}, \citenamefont {Dong},\ and\ \citenamefont
  {Liu}}]{lu2018revealing}%
  \BibitemOpen
  \bibfield  {author} {\bibinfo {author} {\bibfnamefont {C.}~\bibnamefont
  {Lu}}, \bibinfo {author} {\bibfnamefont {B.}~\bibnamefont {Gao}}, \bibinfo
  {author} {\bibfnamefont {H.}~\bibnamefont {Wang}}, \bibinfo {author}
  {\bibfnamefont {W.}~\bibnamefont {Wang}}, \bibinfo {author} {\bibfnamefont
  {S.}~\bibnamefont {Yuan}}, \bibinfo {author} {\bibfnamefont {S.}~\bibnamefont
  {Dong}}, \ and\ \bibinfo {author} {\bibfnamefont {J.-M.}\ \bibnamefont
  {Liu}},\ }\href@noop {} {\bibfield  {journal} {\bibinfo  {journal} {Advanced
  Functional Materials}\ }\textbf {\bibinfo {volume} {28}},\ \bibinfo {pages}
  {1706589} (\bibinfo {year} {2018})}\BibitemShut {NoStop}%
\bibitem [{\citenamefont {Nichols}\ \emph {et~al.}(2016)\citenamefont
  {Nichols}, \citenamefont {Gao}, \citenamefont {Lee}, \citenamefont {Meyer},
  \citenamefont {Freeland}, \citenamefont {Lauter}, \citenamefont {Yi},
  \citenamefont {Liu}, \citenamefont {Haskel}, \citenamefont {Petrie} \emph
  {et~al.}}]{nichols2016emerging}%
  \BibitemOpen
  \bibfield  {author} {\bibinfo {author} {\bibfnamefont {J.}~\bibnamefont
  {Nichols}}, \bibinfo {author} {\bibfnamefont {X.}~\bibnamefont {Gao}},
  \bibinfo {author} {\bibfnamefont {S.}~\bibnamefont {Lee}}, \bibinfo {author}
  {\bibfnamefont {T.}~\bibnamefont {Meyer}}, \bibinfo {author} {\bibfnamefont
  {J.}~\bibnamefont {Freeland}}, \bibinfo {author} {\bibfnamefont
  {V.}~\bibnamefont {Lauter}}, \bibinfo {author} {\bibfnamefont
  {D.}~\bibnamefont {Yi}}, \bibinfo {author} {\bibfnamefont {J.}~\bibnamefont
  {Liu}}, \bibinfo {author} {\bibfnamefont {D.}~\bibnamefont {Haskel}},
  \bibinfo {author} {\bibfnamefont {J.}~\bibnamefont {Petrie}},  \emph
  {et~al.},\ }\href@noop {} {\bibfield  {journal} {\bibinfo  {journal} {Nature
  Communications}\ }\textbf {\bibinfo {volume} {7}} (\bibinfo {year}
  {2016})}\BibitemShut {NoStop}%
\bibitem [{\citenamefont {Yi}\ \emph {et~al.}(2016)\citenamefont {Yi},
  \citenamefont {Liu}, \citenamefont {Hsu}, \citenamefont {Zhang},
  \citenamefont {Choi}, \citenamefont {Kim}, \citenamefont {Chen},
  \citenamefont {Clarkson}, \citenamefont {Serrao}, \citenamefont {Arenholz}
  \emph {et~al.}}]{yi2016atomic}%
  \BibitemOpen
  \bibfield  {author} {\bibinfo {author} {\bibfnamefont {D.}~\bibnamefont
  {Yi}}, \bibinfo {author} {\bibfnamefont {J.}~\bibnamefont {Liu}}, \bibinfo
  {author} {\bibfnamefont {S.-L.}\ \bibnamefont {Hsu}}, \bibinfo {author}
  {\bibfnamefont {L.}~\bibnamefont {Zhang}}, \bibinfo {author} {\bibfnamefont
  {Y.}~\bibnamefont {Choi}}, \bibinfo {author} {\bibfnamefont {J.-W.}\
  \bibnamefont {Kim}}, \bibinfo {author} {\bibfnamefont {Z.}~\bibnamefont
  {Chen}}, \bibinfo {author} {\bibfnamefont {J.}~\bibnamefont {Clarkson}},
  \bibinfo {author} {\bibfnamefont {C.}~\bibnamefont {Serrao}}, \bibinfo
  {author} {\bibfnamefont {E.}~\bibnamefont {Arenholz}},  \emph {et~al.},\
  }\href@noop {} {\bibfield  {journal} {\bibinfo  {journal} {Proceedings of the
  National Academy of Sciences}\ }\textbf {\bibinfo {volume} {113}},\ \bibinfo
  {pages} {6397} (\bibinfo {year} {2016})}\BibitemShut {NoStop}%
\bibitem [{\citenamefont {Matsuno}\ \emph {et~al.}(2016)\citenamefont
  {Matsuno}, \citenamefont {Ogawa}, \citenamefont {Yasuda}, \citenamefont
  {Kagawa}, \citenamefont {Koshibae}, \citenamefont {Nagaosa}, \citenamefont
  {Tokura},\ and\ \citenamefont {Kawasaki}}]{matsuno2016interface}%
  \BibitemOpen
  \bibfield  {author} {\bibinfo {author} {\bibfnamefont {J.}~\bibnamefont
  {Matsuno}}, \bibinfo {author} {\bibfnamefont {N.}~\bibnamefont {Ogawa}},
  \bibinfo {author} {\bibfnamefont {K.}~\bibnamefont {Yasuda}}, \bibinfo
  {author} {\bibfnamefont {F.}~\bibnamefont {Kagawa}}, \bibinfo {author}
  {\bibfnamefont {W.}~\bibnamefont {Koshibae}}, \bibinfo {author}
  {\bibfnamefont {N.}~\bibnamefont {Nagaosa}}, \bibinfo {author} {\bibfnamefont
  {Y.}~\bibnamefont {Tokura}}, \ and\ \bibinfo {author} {\bibfnamefont
  {M.}~\bibnamefont {Kawasaki}},\ }\href@noop {} {\bibfield  {journal}
  {\bibinfo  {journal} {Science Advances}\ }\textbf {\bibinfo {volume} {2}},\
  \bibinfo {pages} {e1600304} (\bibinfo {year} {2016})}\BibitemShut {NoStop}%
\end{thebibliography}%

\newpage\newpage



\section*{Supplementary material}

# Supplementary information for: Anisotropic magnetoresistance in spin-orbit semimetal SrIrO$_3$

D. J. Groenendijk,[*] N. Manca, J. de Bruijckere, A. M. R. V. L. Monteiro, R. Gaudenzi, H. S. J. van der Zant, and A. D. Caviglia

*Kavli Institute of Nanoscience, Delft University of Technology,*
*Lorentzweg 1, 2628 CJ Delft, Netherlands*
Dated: January 20, 2020

## CONTENTS

---

[*] d.j.groenendijk@tudelft.nl

## I. GROWTH OF SIO AND STO THIN FILMS

Figure S1 shows the RHEED intensity during the growth of 5, 6, and 30 u.c. SIO films on $TiO_2$-terminated STO(001) substrates. The growth of SIO is followed by the growth of a 10 u.c. STO film to enable patterning of Hall bars and prevent degradation of the SIO. The clear intensity oscillations indicate that both the SIO and STO grow in layer-by-layer mode.

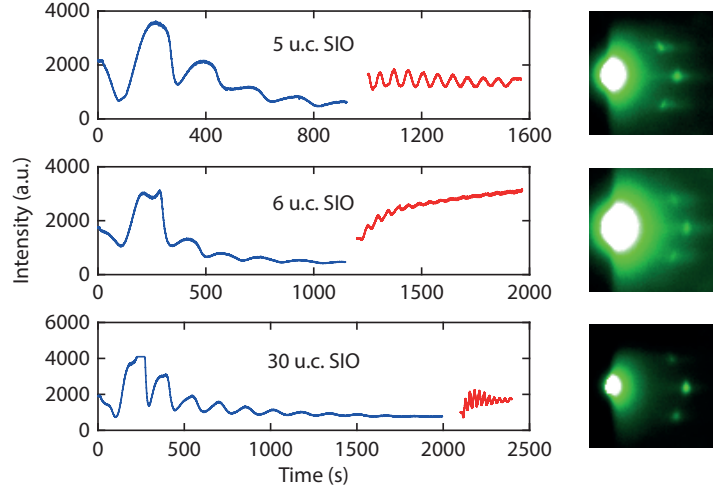

FIG. S1: **RHEED during the growth of SIO and STO.** From top to bottom: RHEED oscillations during the growth of 5 u.c., 6 u.c., and 30 u.c. SIO films, followed by 10 u.c. STO. For the 30 u.c. film, only 12 oscillations are shown. Right: RHEED pattern after the growth of SIO.

## II. MAGNETORESISTANCE OF 5 AND 30 U.C. SIO FILMS

Magnetoresistance (MR) measurements of 30 and 5 u.c. SIO films are shown in Fig. S2. The magnitude of the MR of the 30 u.c. film increases with decreasing temperature (Fig. S2a). Below 4.2 K, the MR displays a cusp at low fields which can be attributed to quantum corrections.

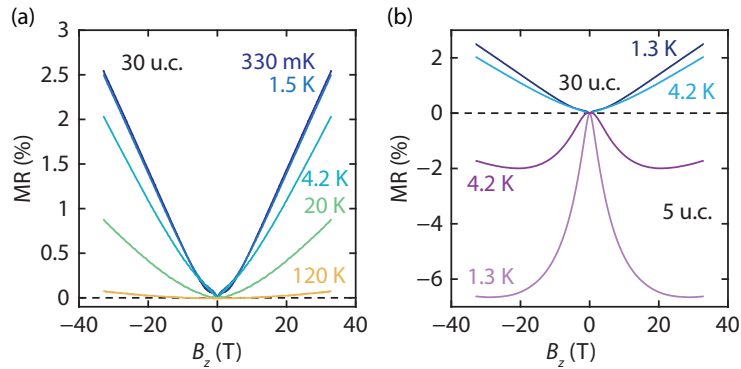

FIG. S2: **Temperature evolution of the MR of 30 and 5 u.c. SIO films.** (a) Temperature-dependent MR of the 30 u.c. SIO film. (b) MR of the 30 and 5 u.c. SIO films measured at 4.2 K and 1.3 K.

Figure S2 shows the MR of the 30 and 5 u.c. SIO films measured at 4.2 K and 1.3 K. The 5 u.c. film displays a large negative MR due to weak localization.

Figure S3a shows the MR of the 30 and 5 u.c. films measured at 1.3 K for $B$ oriented parallel (red) and perpendicular (black) to the film normal.

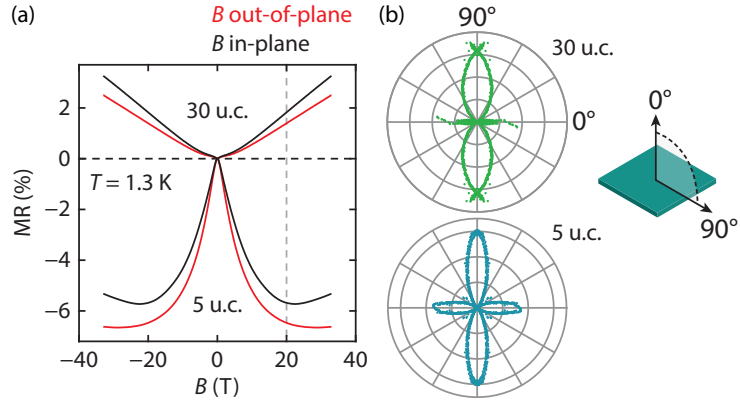

FIG. S3: **Angular dependence of the MR.** (a) MR of the 30 and 5 u.c. SIO films for $B \parallel n$ (red) and $B \perp n$ (black), where $n$ is the film normal. (b) Polar plots of the MR. To the right, a schematic of the two orientations is shown.

For both films, the negative MR is larger when $B$ is oriented in the film plane. To study the anisotropy in more detail, the MR is measured as a function of angle with an applied magnetic field of 20 T. The polar plots in Fig. S3b show two large and two small lobes, similar to the measurements on the 6 u.c. film shown in Fig. 2b (bottom) and 3a (bottom). We attribute the large suppression at $\phi = 90°$ and $\phi = 270°$ to the anisotropy of weak (anti)localization. The other two lobes at $\phi = 0°$ and $\phi = 180°$ arise from the fourfold symmetric signal, which we attributed to magnetocrystalline AMR. This indicates that films of different thicknesses also develop a field-induced magnetization at low temperatures. The AMR appears to become larger as the film thickness is reduced, however its precise evolution should be subject to further investigation.



\end{document}